\documentclass[aps,twocolumn,pre]{revtex4-2}
\usepackage{graphicx}
\usepackage{amsmath,mathtools,esint}
\usepackage{color,ulem,soul}
\usepackage{booktabs}
\usepackage{hyperref}

\newcommand{\re}{\mathrm{Re}}

\begin{document}

\title{Collective Emission in LH2 Assembly Beyond the Point-Dipole Approximation}
\author{Javed Akhtar}
\author{Himangshu Prabal Goswami}
\email{hpg@gauhati.ac.in}
\affiliation{QuAinT Research Group, Department of Chemistry, Gauhati University, Jalukbari, Guwahati-781014, Assam, India}
\date{\today}

\begin{abstract}
Collective emission in light-harvesting assemblies is governed by the local transition dipole and finite geometry of emitting units, a fact that point-dipole approximation obscures. To go beyond this picture, we develop a  non-Hermitian Hamiltonian using the quantum electrodynamic dyadic Green's tensor for a purple bacteria. We construct it for  the isolated 24-bacteriochlorophyll conical frustum and its P42$_1$2 crystallographic assembly. The P42$_1$2 unit-cell symmetry is found to invert the bright-dark ordering of the single ring, placing subradiant states at the low-energy end and revealing the entire crystal to be  the energy-harvesting entity. Tilt-driven switching is activated only in crystal geometries where the finite dipole-carrier (LH2) lies perpendicular to the growth plane. Vacancy and orientational disorder work only in cooperation to renormalize the switching threshold from higher polar angles to lower values.

\end{abstract}

\maketitle

Collective emission in molecular assemblies is usually understood in a Frenkel-exciton basis \cite{Hestand2018, Abramavicius2009, Bustamante2022}, where each chromophore is a localized two-level dipole and the collective response is built from these elementary units\cite{spano1989superradiance}. In this picture, collective behaviour emerges as transition dipoles get arranged with sufficient spatial coherence so that radiative decay gets enhanced in some collective modes and suppressed in others, leading to superradiance and subradiance\cite{Dicke1954,spano1989superradiance,Fidder1991}. Extensive studies of light-harvesting (LH) complexes, which combine high symmetry with structural complexity, established this framework and now motivate the design of controlled collective emission in engineered quantum emitters \cite{Monshouwer1997, zhao1999superradiance,celardo2014cooperative,reitzCooperativeQuantumPhenomena2021}.
The collective radiative response is also shaped by unit-cell symmetry\cite{holman2001,rajasree2022superradiance}, packing anisotropy\cite{celardo2014cooperative,valzelliLargeScaleSimulations2024a,cremer_2020}, and orientational disorder\cite{celardo2014cooperative,valzelliLargeScaleSimulations2024a,cremer_2020}, each of which reshuffles the balance between bright and dark states. In parallel, exact transition-density calculations show that the point-dipole approximation is reliable only at sufficiently large pigment separations \cite{krueger_1998,scholes_1999,madjet_2006}. At shorter distances, finite-size effects and nonsymmetric structural effects start to modify the optical responses \cite{scholesMechanismLightHarvesting2000,scholesLongRangeResonance2003}. Such multiple concepts, whose effects are well understood in isolation, the combined effect offers a direction for further research esp. when the assembly is not a point dipole but a finite, anisotropic dipole carrier. The idea gets exotic in densely packed systems, where the spatial structure of the assembly and its repetition across the lattice can alter the pathways of constructive and destructive interference, a physical regime less understood although explored\cite{Bustamante2022,cremer_2020,lee_2023}. The LH complexes provide a natural platform to address this issue at the fundamental level. For example, the LH2 complex of the purple bacterium \textit{Rhodospirillum molischianum} consists of bacteriochlorophylls (BChls) arranged in a low symmetric conical-frustum geometry, with B800 and B850 pigments adopting distinct orientations relative to the membrane plane in the P42$_1$2 unit cell \cite{koepkeCrystalStructureLightharvesting1996a}. In this case, the conical frustum should be viewed as a finite, structurally anisotropic dipole carrier rather than a featureless point object embedded in an overall morphologically flat crystal. The analogy is closely related to crystal-engineering platforms, where unit-cell motif geometry and molecular orientation govern excitonic coupling and emission in low-dimensional molecular crystals and co-crystals \cite{haldar2019novo,kim2023plane,guerrini2018solid,deka2024exceptional}.

Treating finite geometry on the same footing as the electromagnetic field is possible for ordered arrays and large biological assemblies via the macroscopic quantum electrodynamics formulation involving the dyadic Green tensor \cite{holzinger2022cooperative,palEfficientExcitationTransfer2025,celardoSuperradiancePhotosynthetic2012,valzelliLargeScaleSimulations2024a}. The coherent interaction and collective radiative decay arise together from the Green tensor, so the effective Hamiltonian is non-Hermitian by construction rather than by phenomenological addition of loss\cite{lehmbergRadiationNAtomSystem1970,lehmberg1970radiation}. The eigenmodes are then the radiative analogues of collective excitons \cite{Monshouwer1997}. The formalism allows the full crystallographic symmetry, spatial information, orientation, retardation and isolated decays  through the coherent and dissipative couplings, when expressed in an engineered basis. In this sense, the dyadic Green formalism goes beyond the standard point-dipole and Frenkel descriptions by providing a geometry-resolved molecular quantum-electrodynamic framework. For crystals such as LH2, this opens a route beyond the usual local-exciton picture by explicitly incorporating finite dipole carriers that are hierarchically collective across multiple length scales.  The present work differs in that the chromophoric unit is not treated as a point emitter. Instead, the BChl geometry enters directly through the spatial and orientational dependence in the dyadic Green function, thereby going beyond the dipole approximation.

 Our study  begins with the full 24-BChl LH2 which we hierachially  assemble into crystallographic P42$_1$2 unit cell and finally into anisotropic finite flat slabs from a quantum electrodynamic perspective. We construct the radiative collective emission spectrum of the crystal by retaining the conical frustum shape and dipole orientational information of the isolated LH2 (dipole carrier).  We observe a bright to dark switching behaviour which depends not only on the dipole orientation but is sensitive to the geometrical arrangement of LH2 in the crystal. Whenever the dipole carrier is parallel while the effective dipole orientation is perpendicular to the crystal's growth plane, a switch between low energy subradiant states and high energy superradiant states is observed on rotating the structured dipole. The orientational space is robust in presence of vacancy as well as configurational disorder. Vacancies and disorder can actually assist  by reducing the geometric reorganization.

\begin{figure}
    \centering
    \includegraphics[width=\linewidth]{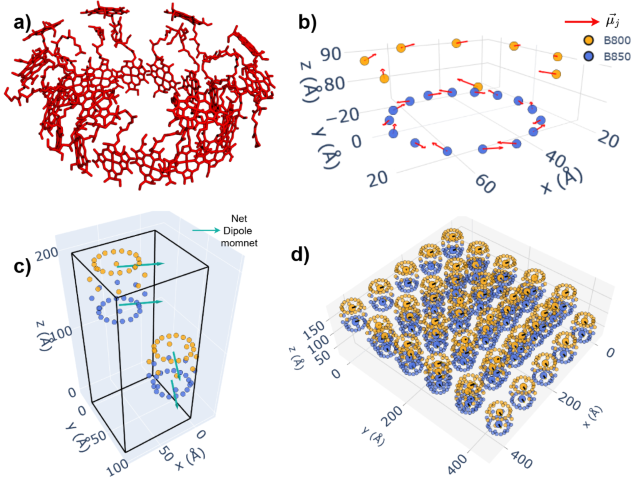}
    \caption{(a) A space-filling representation of the arrangement of the 24 bacteriochlorophyll (BChl) chromophores within the LH2 protein complex. (b) The simplified physical model of a single LH2 ring, where each of the $N_{\text{pig}} = 24$ BChl chromophores is treated as a two-level system. The orientation of each pigment's transition dipole moment ($Q_y$) is indicated by a red arrow. (c) The crystallographic P42$_1$2 unit cell, showing the symmetric arrangement of four identical LH2 "super-atom" units whose dipole orientations are set by symmetry. (d) Macroscopic LH2 antenna assembly obtained by tiling the $\mathrm{P4_2 2_1 2}$ unit cell along all three crystallographic axes}
    \label{fig-1}
\end{figure}

At the microscopic level, an LH2 complex (Fig. (\ref{fig-1})a) of Rh. molischianum is  a 24-pigment excitonic bacteriochlorophyll  aggregate comprising of B800 (8 BChl units) and B850 (16 BChl units) subrings. Each BChl unit that contribute to the rings is described as an effective exciton \cite{koepkeCrystalStructureLightharvesting1996a} with a fixed position (taken from X-ray resolved PDB databank)  as shown in Fig. (\ref{fig-1}b), a geometry mimicking the frustum of a cone. Each exciton is initially placed at the centre of mass of the bacteriochlorophyll unit that make up a single LH2 complex. 
The dynamics of the single LH2 complex (Fig. (\ref{fig-1}b)) follows an effective exciton-basis dependent non-Hermitian Hamiltonian that captures both coherent exciton transfer and radiative decay \cite{lehmbergRadiationNAtomSystem1970}. It is of the form, $H_{\mathrm{eff}}^{\mathrm{LH2}} = \hat H_{\downarrow} +\hat H_{\uparrow}+ \hat H_{\updownarrow}$ (details in SI), representing the sum of lower (B850), upper (B800) and inter-ring contributions to the energetics of the single LH2 complex. The intra- and inter-ring couplings are evaluated from a dyadic Green's tensor \cite{Dung2002,agarwal_1970} , expressed in the exciton basis and depend on the two different excitation frequencies (800 nm and 850 nm), detuning between the two, spatial pigment separation and dipole orientation (see SI for the exact Hamiltonian).  The orientation (exact head to tail arrangement) and magnitude of the dipole moment used in the Green's tensor is constructed from the Q$_y$ transition-dipole orientation of a single BChl \cite{oviedo2011transition}. The B800 and B850 subunits get distinguished by their site energies and decay scales in the Green's tensor.

Multiple LH2 complexes are known to get arranged in P42$_1$2 space-group forming LH2 crystals \cite{koepkeCrystalStructureLightharvesting1996a,sundstrom1999photosynthetic}. To describe a large LH2 crystal, we  first create a geometrically accurate  P42$_1$2 unit cell  with four single LH2 complexes per unit cell (Fig. (\ref{fig-1}(c)) with data from established spatial and orientational information \cite{sundstrom1999photosynthetic},  which we call the LH2 unit cell.  Repeating the unit cell in different directions of the three axes leads to a crystallographically consistent representation of the overall macroscopic version of the LH2 antenna assembly as shown in Fig. (\ref{fig-1}d).  Through this hierarchial approach, we not only provide a quantum electrodynamic modeling of excitons in LH2, but also  use it to unravel  the hierarchy of the radiative collective spectrum from the bare pigment to the full crystal. We achieve this by taking into account the dipolar physics at two levels of Green's tensor based non-Hermitian dynamics. The first level resolves the radiative structure of the 24-pigment LH2 ring (converts site dipoles into molecular collective modes with approriate infomation from the position vectors). The second level translates the relevant excitons from the earlier level to an effective LH2 crystal (converts molecular collective modes into crystal collective modes). The novelty is therefore not a new microscopic multipole expansion, but a symmetry-resolved, geometry-aware quantum-electrodynamic reduction from pigments to  crystals. That is closer to a renormalization-style change of basis rather than a repetition. 
\begin{figure}
    \centering
    \includegraphics[width=1\linewidth]{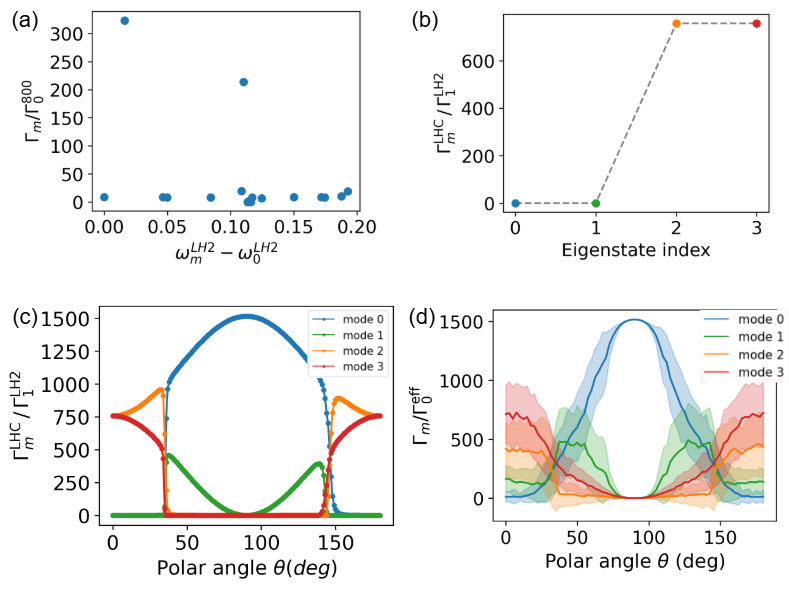}
    \caption{(a) Scaled decay-rate spectrum $\Gamma_m/\Gamma_0^{B800}$ of a single isolated LH2 ring, plotted against the eigenfrequency offset $\omega_m^{LH2} - \omega_0^{LH2}$ from the ground state; the dominant superradiant state is concentrated near the low-energy end of the excitonic manifold, while the majority of states are subradiant across the full spectral range, 
    (b) Decay-rate spectrum of the crystallographic unit cell comprising four LH2 super-atoms
    (c) Continuous tilt-angle dependence of all four eigenmode decay rates of unit cell, a dark-to-bright inversion of the excitonic ground state is identified at the crossing angles $\theta\approx 35^\circ$ and $145^\circ$, with the Dicke superradiant plateau sustained over $\theta\in[45^\circ,135^\circ]$. 
    (d) Ensemble-averaged decay rates $\pm 1\sigma$ (shaded) over $N=200$ realizations with independently randomized azimuthal orientations $\phi_i \in [0^\circ, 360^\circ)$ per site; mode 0 retains the bell-shaped superradiant enhancement centered at $\theta=90^\circ$ with narrow variance, while modes 1-3 acquire finite mean brightness with large variance}
    \label{fig:tilt}
\end{figure}

The quantum electrodynamic model of this P42$_1$2 unit cell is constructed from the eigenspectrum of a single LH2 complex (i.e post-diagonalisation of $H_{\mathrm{eff}}^{\mathrm{LH2}}$). It leads to 24 eigenstates with eigenvalues $\lambda_m = \hbar{\omega}_m^{LH2}-i{\Gamma}_m^{LH2}/2, m = 0\ldots 23$. The imaginary part of the eigenvalues representing the decay rates of each of the 24 eigenstates, scaled by the decay rate of a single B800-BChl is shown in Fig. (\ref{fig:tilt}a) in ascending eigenenergies. The first two excited states and the 8th and 9th states are superradiant while the rest are subradiant. The association of the terminology is also confirmed by evaluating the near-field and far field emission profiles (see the SI), where we observe distinct  distributions for these four eigenstates. This distinct nature of the two different collective emission profiles in the low and mid end of the eigenspectrum  is a consequence of the geometry of the LH2 frustum arrangement. There are four unique configurationally different types of spatial and orientational dipole-pairs in the two rings of the LH2 frustum (fig \ref{fig-1} b). Upper ring has a head-tail configuration, lower ring has head-head or tail-tail configurations. Inter-ring configrations can be head(tail)-head(tail) or head-tail realisations. A single circular array with a single dipole orientation is known to generate a single superradiant peak \cite{spano1989superradiance, moreno-cardonerSubradianceenhancedExcitationTransfer2019}. Thus, four possible configurations lead to four superradiant states.  

 To construct an effective non-Hermitian Hamiltonian for the unit cell, we restrict the energy manifold to the ground state $|G\rangle$ (no excitation on the LH2 unit) and its brightest super-radiant eigenstate $|S\rangle$ of the single LH2.  This reduction is justified by the assumption that, under the low-intensity excitation conditions relevant for natural photosynthesis, the system dynamics are dominated by a small subset of the 24 excitonic eigenstates. The brightest super-radiant eigenstate couples most strongly to the far-field radiation \cite{meier1997polarons}, and can be taken to be the primary optical response of the complex, thereby mapping each LH2 complex onto an effective collection of  two-level systems, akin to super-atom (SA) picture commonly used in collective and cooperative quantum optical systems \cite{kumlin2023quantum}.  The effective Hamiltonian of the LH2 crystal in the eigenbasis of a single LH2 (rotating with respect to the difference in frequency of the eigenstates $|G\rangle$ and $|S\rangle$, i.e $\omega_{1}^{LH2}-\omega_{0}^{LH2}=\omega_0^{\mathrm{eff}}$ ), arranged  in the P42$_1$2 lattice takes the form,
\begin{align}
H_{\mathrm{eff}}^{\mathrm{LHC}}=
\hbar\sum_{k,l}^{4N}
\Bigl(\Omega_{kl}^{\mathrm{LH}}-\tfrac{i}{2}\Gamma_{kl}^{\mathrm{LH}}\Bigr)
\hat\tau_k^{\dagger}\hat\tau_l^{}. 
\label{eq:Heff_lattice}
\end{align}
$\hat{\tau}_k = |G\rangle\langle S|_k$ is the effective de-excitation operator for the $k$th LH2 frustum and $\hat{\tau}_k^{\dagger} = |S\rangle\langle G|_k$ is its adjoint and $N$ denotes the number of P42$_1$2 unit cells, with four super-atoms per cell. The coherent coupling $\Omega_{kl}^{\mathrm{LH}}$ and dissipative coupling $\Gamma_{kl}^{\mathrm{LH}}$ between the $k$th and $l$th super-atoms are obtained using appropriately re-calibrated dyadic Green-tensor formalism as at the pigment level [SI, Eqs.~(S3), (S4)], evaluated at the single-SA transition 
frequency $\omega_0^{\mathrm{eff}}$, defined as the energy difference between $|S\rangle$ and $|G\rangle$ in the eigenspectrum of $H_{\mathrm{eff}}^{\mathrm{LH2}}$. The renormalized transition dipole orientation used in the Green's tensor  is taken to be a coherent vector sum of the 24 contributing dipoles using the eigenvector $|S\rangle$:
\begin{align}
  \boldsymbol{\mu}_\mathrm{SA} =
  \sum_{j=1}^{24} c_j^{(S)}\,\re \{\boldsymbol{\mu}_j\},
  \label{eq:musa}
\end{align}
where $c_j^{(S)}$ is the $j$th component of the eigenstate $|S\rangle$. The resultant dipoles are no longer preserve the original LH2 frustum's dipole arrangement (Fig. 1c). Diagonalization of Eq.~(\ref{eq:Heff_lattice}) yields collective crystal eigenmodes of the form $\zeta_m=\hbar{\omega}_m^{LHC}-i{\Gamma}_m^{LHC}/2$.  The superradiant and subradiant crystal eigenstates are labelled according to whether their radiative widths are enhanced or suppressed relative to the single LH2 respectively. 

\begin{figure}
    \centering
    \includegraphics[width=\linewidth]{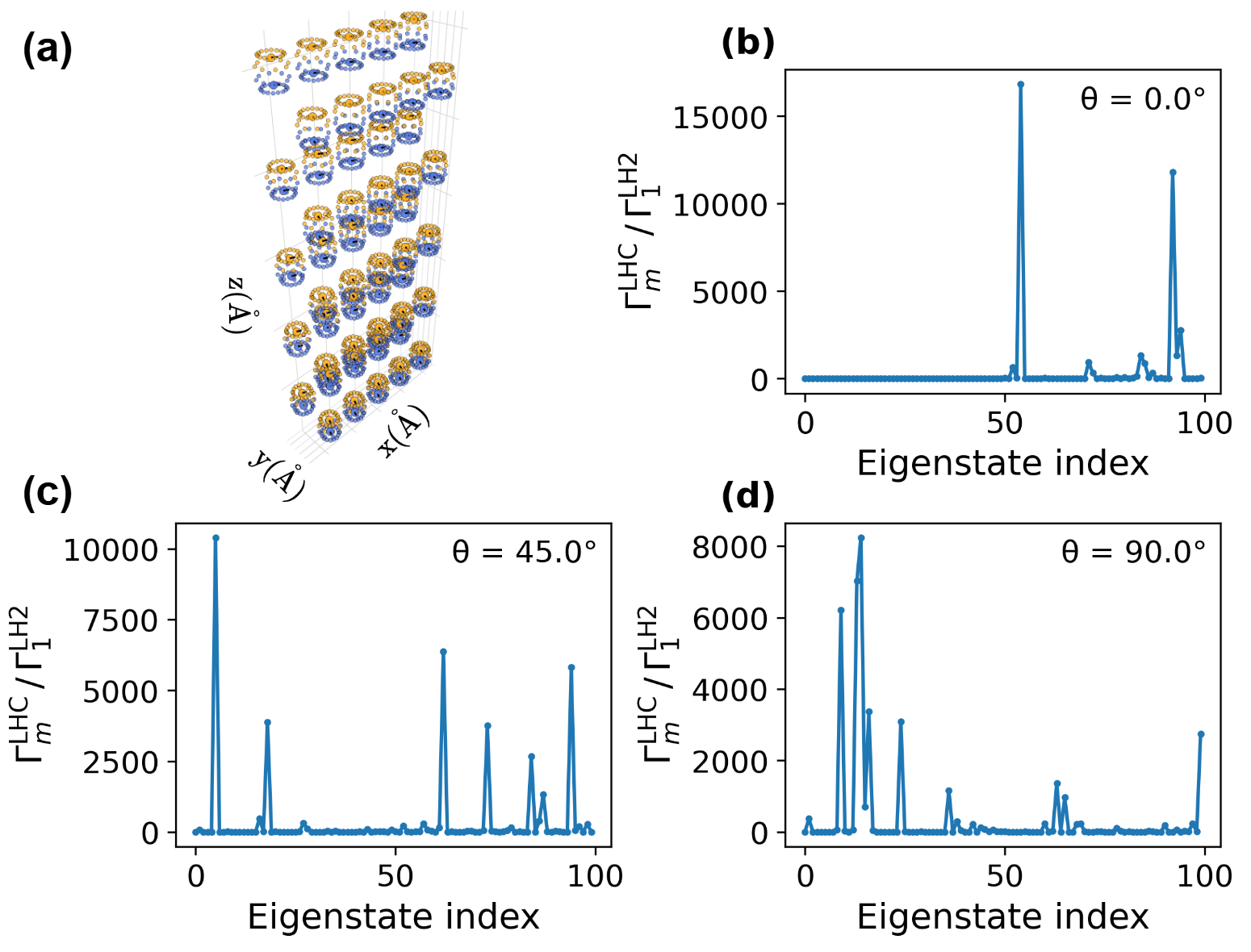}
    \caption{Scalled collective radiative decay-rate spectra of a pristine $5\times1\times5$
    LH2 crystal slab, with super-atom dipoles tilted by $\theta$ about the $x$-axis.
(a) Three-dimensional visualization; blue and orange complexes denote  upward- and downward-facing LH2 super-atoms related by the $\mathrm{P42_12}$ flip operation. (b--d) Scaled decay-rate spectra at $\theta = 0^\circ$, $45^\circ$, and $90^\circ$ respectively. At $\theta = 0^\circ$ (b), two sharp superradiant peaks appear near eigenstate indices 60 and 95 ($\Gamma_{\max}/\Gamma_0^{\mathrm{eff}} \approx 16{,}500$). At $\theta = 45^\circ$ (c), superradiant weight migrates to low eigenstate indices (0--25), $\Gamma_{\max}/\Gamma_0^{\mathrm{eff}} \approx 10{,}500$. At $\theta = 90^\circ$ (d), superradiant modes concentrate at the lowest eigenstate indices, $\Gamma_{\max}/\Gamma_0^{\mathrm{eff}} \approx 8{,}000$. confirming the tilt-driven inversion of the excitonic ground state}
    \label{fig_3_515}
\end{figure}

\begin{figure}
    \centering
    \includegraphics[width=\linewidth]{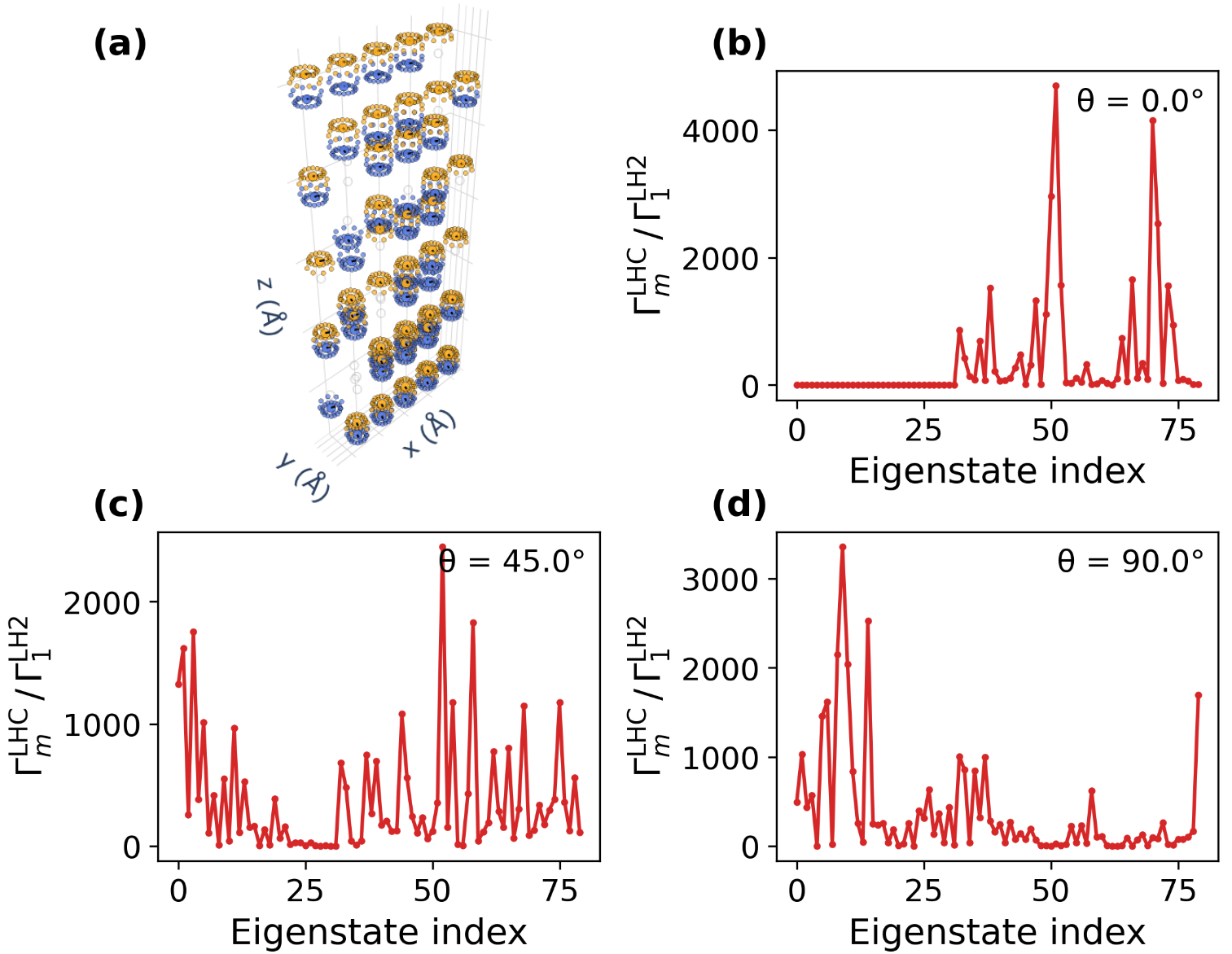}
    \caption{Scaled decay-rate spectra of an 80-superatom $5\times1\times5$ LH2 crystal slab with 20\% random site vacancy, with super-atom dipoles tilted by $\theta$ about the $x$-axis. (a) Three-dimensional visualization of the vacancy-disordered rod; blue and orange complexes denote upward- and downward-facing LH2 super-atoms respectively. (b--d) Decay-rate spectra at $\theta = 0^\circ$, $45^\circ$, and $90^\circ$ respectively. At $\theta = 45^\circ$ (c), oscillator strength is distributed broadly across all eigenstate indices ($\Gamma_{\max}/\Gamma_0^{\mathrm{eff}} \approx 2{,}500$). At $\theta = 90^\circ$ (d), low-index superradiant modes are recovered with $\Gamma_{\max}/\Gamma_0^{\mathrm{eff}} \approx 3{,}500$.}
    \label{fig515_20}
\end{figure}

For a single P42$_1$2 unit cell, N = 1 and  Eq. (\ref{eq:Heff_lattice}) yields four complex eigenmodes. The scaled decay spectrum for the four modes of the unit cell is shown in Fig.~\ref{fig:tilt}(b). The four unit-cell modes have two dark states (modes 0 and 1) and two  degenerate super-radiant states (modes 2 and 3). In contrast to the single LH2, the bright states now occupy the top of the energy manifold. The P42$_1$2 symmetry therefore inverts the energetic ordering established at the single-LH2 level. The excitations are effectively trapped directly resulting in long lived energy, i.e harvesting. So, it is the unit cell that leads to harvesting properties rather than a single LH2 unit. Thus, energy harvesting is a collective phenomenon, driven by collective dissipative couplings between contributing LH2 units that promote transfer of excitation from high energy superradiant states to low energy subradiant states. The switching of low energy superradiant states to low energy subradiant states as the heirarchy is changed from single LH2 to a light harvesting crystal (LHC) is a direct result of the change  of the conical frustum symmetry with head-tail arrangement of the LH2's dipole moments  to the crystallographic symmetry with a different orientation of the four effective transition dipole moments of LHC unit cell.
To understand the effect of dipole orientation on the decay rate, 
we can rotate the  effective dipole of the super-excitons of the contributing LH2 units through the polar angle $\theta$, outward from the $xy$-plane configuration ($\theta=0^\circ$). The angle serves as an auxilliary control parameter, $\theta \in \{0, 90^\circ\}$.  The resulting evolution of the four decay modes is displayed in Fig.~\ref{fig:tilt}(c) and is symmetric about $\theta=90^\circ$. The eigenstates are tracked across the polar  sweep by an eigenvector overlap-based assignment to preserve the mode identity through avoided crossings. As $\theta$ increases, the decay rate degeneracy of the two high energy bright modes get lifted. One mode becomes increasingly radiative, while the other is progressively suppressed. At $\theta=45^\circ$,  the eigenstate character is redistributed through an avoided crossing resulting in  switch between bright and dark modes. The low energy dark modes become low energy bright modes and vice versa.   At $\theta=90^\circ$, all dipoles align along $\hat{z}$ and a single super-radiant mode emerges at the bottom of the manifold, while the remaining three modes become dark. This trend complements itself as the polar angle moves from 90 to 180.  
At $\theta=0^\circ$, the bright state lies above the dark manifold, whereas at $\theta=90^\circ$ it becomes the lowest-lying mode.  
To test the robustness of the bright-dark switching, we introduce an orientational disorder by randomizing the azimuthal angle of each super-exciton dipole while keeping the polar angle $\theta$ fixed. This type of disorder introduces a configurational defect where the unit cell remains fixed but the orientation of the LH2 units change in the azimuthal plane. As shown in Fig.~\ref{fig:tilt}(d), for different polar angles, the Monte-Carlo based ensemble-averaged decay spectrum becomes strongly asymmetric in decay-rate magnitude but symmetric across the polar sweep. The bright-dark switching behaviour happens gradually but remains robust. However, the polar angle at which the switching occurs is lowered in comparsion to the case without disorder. The first two dark and bright (mode 0 and 2) switching happens at $\theta \approx 36^\circ$.  The key effect of disorder is not to remove the bright-dark crossover, but to shift its onset. The critical polar angle for the appearance of superradiance is reduced in the disorder-averaged case. Thus, azimuthal disorder lowers the geometric threshold for cooperative brightening, indicating that the switch is robust but renormalized by  orientational fluctuations. In this sense, azimuthal disorder relaxes the cancellation pathways that suppress collective emission, acts cooperatively with polar orientation, enabling superradiance at smaller out-of-plane angles. We attribute this observation to a thermodynamic reading: azimuthal disorder increases the configurational freedom of the dipolar array, while the collective bright state is favoured when the orientational constraints are sufficiently relaxed to overcome destructive interference. In free-energy language, the superradiant state is stabilized by an enthalpy-like cooperative alignment of dipoles, whereas the in-plane orientational disorder contributes a configurational entropy term. The observed switch occurs when the balance tips in favour of the bright collective mode. Azimuthal disorder increases the configurational entropy of the dipolar ensemble while reducing the orientational constraint needed to assemble a cooperative radiative state. The bright–dark inversion can then be viewed as an enthalpy-like stabilization of the superradiant configuration, reached at a smaller polar angle because the in-plane disorder relaxes destructive-interference pathways and lowers the effective reorganization cost. That interpretation is consistent with the fact that cooperative emission in molecular emitters is strongly controlled by dipole orientation and spatial arrangement, and that configurational entropy is the natural thermodynamic measure of the number of accessible orientational arrangements.

\begin{figure}
       \centering
    \includegraphics[width=\linewidth]{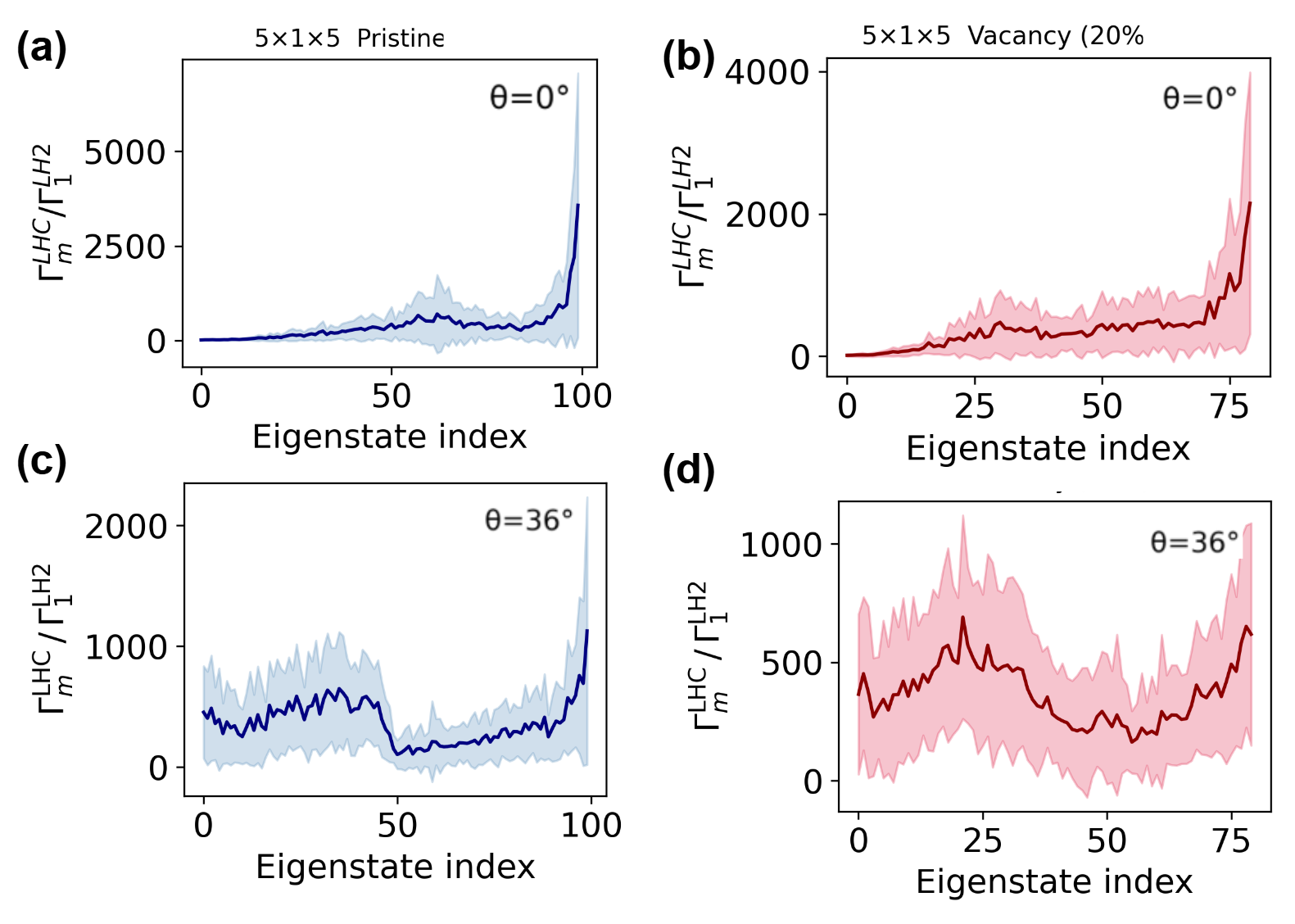}
    \caption{Ensemble-averaged $\Gamma_m^{\text{LHC}}/\Gamma_1^{\text{LH2}}$ spectra for a $5\times1\times5$ LH2 crystal rod ($N=200$ realizations, random $\phi_i$ per site, fixed $\theta$). Solid lines (shading) denote the mean ($\pm 1\sigma$). (a) Pristine ($\theta = 0^\circ$): Superradiant peak at index $\approx 95$ ($\Gamma_{\max} \approx 6000$). Narrow variance suggests the bright doublet is robust against azimuthal randomization. (b) 20\% Vacancy ($\theta = 0^\circ$): Peak persists ($\Gamma_{\max} \approx 4000$) with increased high-index variance; low-index modes remain dark. (c) Pristine ($\theta = 36^\circ$): Spectral weight redistributes across indices 0–50. Redistribution occurs earlier than in $5\times5\times1$ slabs due to narrow $b$-axis excitonic bandwidth. (d) 20\% Vacancy ($\theta = 36^\circ$): Broad profile is preserved ($\Gamma_{\max} \approx 1000$); large variance reflects combined vacancy and azimuthal disorder effects on $a,c$-axis coupling.}
    \label{fig:azim_cry_515}
\end{figure}

The model is next scaled to 100 LH2 units in a two-dimensional planar geometry, mimicking the natural flat morphology of the LH2 assembly~\cite{huPhotosyntheticApparatusPurple2002,kulkarniTheoreticalStudyInfluence2024,koepkeCrystalStructureLightharvesting1996a, hsin2010self}. Two slab morphologies are possible, transverse and longitudinal. We construct a  transverse slab (5 × 1 × 5), where five unit cells are tiled along a and five along c, with a single unit cell along b. The growth plane is ac. In the longitudinal slab (5 × 5 × 1), five unit cells are tiled along a and five along b, with a single unit cell along c. The growth plane is ab. 
Both the slabs differ in how the effective vector sum of the individual coherent  dipoles of the four contributing LH2s in the $P42_12$ symmetry is oriented relative to the growth plane.   For the transverse slab, the net dipole carrier (the unit cell) orientation is along the growth plane (ac) and for longitudinal slab it is perpendicular to  the growth plane (ab).   Effectively, these slabs are analogous to  J and H-aggregates~\cite{Hestand2018, Spano2010}. For the longitudinal case, each effective neighbouring dipoles are found to be aligned head-to-tail along the stacking direction resembling a J -aggregate. The transverse slab resembles an H-aggregate, where dipoles are found to finally align side-by-side such that the dipoles potrude out of the growth plane.
At $\theta = 0^\circ$, for both slabs net dipole moment orientation ($ab$-plane) perpendicular to dipole carrier orientation (exhibit H-aggregate  character). And the  superradiant weight concentrated at  the high-energy end of the manifold. As $\theta$ increases, all super-atom dipoles are  tilted uniformly out of the $ab$-plane, acquiring a  $z$-component parallel to the crystallographic  $\mathbf{c}$-axis. In the transverse slab, $\mathbf{c}$ is a dominant  coupling direction; the growing $z$-component is  head-to-tail with respect to it, and J-aggregate coupling along the stacking axis is progressively  activated. In the longitudinal slab, the dominant coupling  channels run along $\mathbf{a}$ and $\mathbf{b}$;  the $z$-component is perpendicular to both, and  H-aggregate character is preserved throughout.

The scaled decay-rate spectrum  $\Gamma_m^{\mathrm{LHC}}/\Gamma_1^{\mathrm{LH2}}$  of the transverse slab is shown in  Fig.~\ref{fig_3_515}(b--d) at $\theta = 0^\circ$,  $45^\circ$, and $90^\circ$. At $\theta = 0^\circ$, two sharp superradiant peaks  are observed at eigenstate indices 60 and 95, with  $\Gamma_{\mathrm{max}}/\Gamma_1^{\mathrm{LH2}}  \approx 16{,}500$. The two peaks arise from independent coherent coupling  channels along $\mathbf{a}$ and $\mathbf{c}$. The low-energy sector (indices 0--40) remains  subradiant (same as bright-above-dark  ordering like the unit cell). At $\theta = 45^\circ$, spectral weight migrates  toward low eigenstate indices (0--25), with  $\Gamma_{\mathrm{max}}/\Gamma_1^{\mathrm{LH2}}  \approx 10{,}500$. At $\theta = 90^\circ$, all net dipoles are directed  along $\hat{z}$ and superradiant modes are  concentrated at the lowest eigenstate indices,  with $\Gamma_{\mathrm{max}}/\Gamma_1^{\mathrm{LH2}}  \approx 8{,}000$. 
The switching is not observed in the longitudinal  slab (SI fig(\ref{fig:S_551_pristine})). There the bright states remain concentrated at high energy and over the same range of angles. The crystal there extends along $\mathbf{a}$ and $\mathbf{b}$, not along $\mathbf{c}$. As $\theta$ increases, the growing $z$-component of the dipoles is perpendicular to the $\mathbf{a}$ and $\mathbf{b}$ neighbours. Side-by-side coupling is maintained, and superradiant weight is retained at the high-energy end at all angles.
The comparison identifies a  condition for tilt-driven switching: the crystal must extend along $\mathbf{c}$, so that the growing $z$-component becomes head-to-tail with respect to the crystal neighbours, driving superradiant weight to the low-energy end.
The natural flat membrane of  \textit{Rh.~molischianum} corresponds to the  longitudinal slab morphology. The LH2 conical frustums stand perpendicular  to the membrane surface. The net dipoles lie within it. The bright-above-dark ordering is preserved at  all realistic dipole orientations. The low-energy subradiant states form a robust energy trapping reservoir. Tilt-driven switching is geometrically suppressed  by the perpendicular placement of the dipole carrier. An individual LH2 frustum does not constitute  the harvesting unit. The harvesting is an emergent property of the  crystal architecture, conditioned on the orientation  of the finite dipole carrier relative to the  growth plane.

The polar angle,  although not a strict natural control-knob,  allows identification of  geometry based finite dipole effects. It also has a practical application. Through this angle, a leverage over artificial quantum systems in controlling the collective emission is possible, not only through the dipole orientation but also through the dipole carrier's spatial orientation.  
We now introduce 20\% random site vacancy in the slabs. This reduces the number of active LH2 units from 100 to 80. The resulting decay spectra (Fig.~\ref{fig515_20} b) for the transverse slab shows no qualitative change at $\theta = 0^\circ$. The subradiant states remain at the low-energy end of the spectrum, although their magnitudes are reduced. As the polar angle increases, the switching over to low-energy superradiant states is retained. However, in the presence of vacancy, this effect is enhanced. At $\theta = 45^\circ$, a larger number of superradiant modes appear at the low-energy end compared to the pristine slab (Fig. \ref{fig515_20}(c)). Thus, site vacancies do not suppress the switch in transverse crystals. Instead, it redistributes and amplifies the low-energy bright sector.  In contrast, the longitudinal geometry's response to polar variation remains intact (Fig.~\ref{fig:S_551_vac} of SI). Vacancies do not promote any switching in the longitudinal LH2. The subradiant states remain concentrated at the low-energy end across all polar angles. Only a broadening of the spectrum is observed. These results establish a clear conclusion. Vacancy alone does not induce any switching. It only promotes the effect when the dipole orientation is already compatible with the underlying crystal's spatial geometry.

Introducing azimuthal disorder does not change the basic picture. The bright states still stay near the high-energy side of the spectrum in the pristine crystal, Fig \ref{fig:azim_cry_551}. With 20\% vacancy, this bright sector remains, but becomes broader and less sharply localized with the lower end being subradiant.  At $\theta \approx 36^\circ$, the bright states shift to lower energy, but they still remain mainly in the middle of the spectrum. Vacancy disorder spreads the bright weight over more eigenstates. The low-energy states remain mostly subradiant. Even, in the presence of disorder, the longitudinal LH crystal's harvesting nature remains robust, as what is observed in nature. The transverse geometry's collective emission behaves the same way as in the absence of disorder, but the switch occurs earlier (Fig~\ref{fig:azim_cry_515}(c)). Bright-dark switching is still present as the polar angle increases. However, vacancy disorder moves the onset to smaller angles. In the disordered case, superradiant states appear  at $\theta \approx 36^\circ$. Thus, disorder does not create a new mechanism. It only lowers the orientational threshold for collective brightening. This strengthens the applicability. In designing artificial harvesters mimicking the natural LH2, the dipole itself is not the only important variable. The geometry that carries the dipole also matters and is robust against vacancy or disorders. The emission is controlled by both dipole orientation and crystal geometry. Vacancies help the system reach the bright regime with less geometric change. This observation can be borrowed to the artificial platforms where cooperative response can be now easier to tune. The geometry of the dipole carrier, together with the dipole orientation itself, becomes an additional control variable for collective emission. The cooperative radiative response is therefore governed not only by the local dipole moment, but also by how that dipole is embedded in a finite anisotropic structure. Natural systems exploit this coupled control of dipole and geometry to tune emission pathways or select a pathway.  The same polar angle tilt acts differently depending  on the crystal geometry. In the longitudinal slab, the growing  $z$-component is perpendicular to the crystal  neighbours; the coupling character does not change. In the transverse slab, the growing  $z$-component is parallel to the $\mathbf{c}$-axis  neighbours; side-by-side coupling is converted  to head-to-tail and switching is activated. 
The subradiant modes are hence supported in the longitudinal crystal. So, there is a preference shown by the bacteria for surface-parallel anisotropic dipole alignment with out of plane structural alignment. The dipole tilt is necessary but not sufficient; the crystal geometry determines whether its lower energy modes will be superradiant or not. Superradiance is possible only when the preference over such a rigid dipole-structure orientational space is relaxed by choosing an alternate spatial orientation and not just the dipole. Only then, the polar angle can be used as a control knob to switch between bright and dark states.

To conclude, in this work we treated the light harvesting  photosynthetic antenna complex of a specific purple bacteria, \textit{Rh. molischianum},  within the framework of quantum electrodynamics by translating  the  finite sized-conical frustrum geometry of the bacteriochlorophylls to its actual experimentally realised crsytallographic symmetry. By constructing a non-Hermitian Hamiltonian and accounting for all LH2 parameters from realistic values, the collective light-induced  dynamics was explored by explicitly evaluating a position, size and dipole dependent  dyadic Green's tensor.   The radiative structure of the bacteria's 24-chromophore conical frustum is first resolved at the pigment level. The resulting brightest collective eigenstate is then carried forward as an effective super-atom, in this case: super-exciton, into the P42$_1$2 crystal Hamiltonian. We found that the emission and energy harvesting is an emergent consequence of its crystallographic arrangement rather than from the isolated LH2 moeity.   The low-energy subradiant manifold of the crystal is what constitutes the energy-trapping entity associated with harvesting. The polar tilt angle of the super-exciton dipoles is found to continuously redistribute collective decay rates across the eigenspectrum. A geometry-conditioned bright-to-dark inversion is identified, whose onset is controlled by whether the growing out-of-plane dipole component is head-to-tail or side-by-side with respect to the crystal neighbours. The unit-cell symmetry alone suffices to invert the bright-above-dark ordering of the single LH2 moiety. Vacancy disorder and azimuthal orientational randomness are found to renormalize, rather than destroy, this switching behaviour. The natural membrane morphology of the bacteria, corresponding to a longitudinal slab geometry, is found to geometrically suppress the switching, preserving the low-energy dark manifold at all physically accessible dipole orientations. The harvesting function is therefore concluded to be an emergent property of the crystal architecture. Both the dipole and its structural carrier act as jointly necessary determinants of the collective radiative response, a conclusion whose generality is applicable across broader classes of biological and reverse-engineered artificial antenna or antenna assemblies.

\section*{Supplementary Information}

\appendix

\renewcommand{\theequation}{S\arabic{equation}}
\renewcommand{\thefigure}{S\arabic{figure}}
\renewcommand{\thetable}{S\arabic{table}}
\setcounter{equation}{0}
\setcounter{figure}{0}
\setcounter{table}{0}
\section*{S1.\;Effective Hamiltonian of the Single LH2 Complex}
 \label{sec:S1}

The LH2 complex from \textit{Rhodospirillum molischianum} is partitioned
into an eight-chromophore B800 sub-ring and a sixteen-chromophore B850
sub-ring in the site basis provided by the crystallographic structure of
Koepke et al.~\cite{koepkeCrystalStructureLightharvesting1996a}.
Restricting to the single-excitation manifold and adopting the
Born--Markov approximation, the reduced dynamics of the
$N_\mathrm{pig}=24$ two-level emitters are generated by the
non-Hermitian effective Hamiltonian\cite{holzinger2022cooperative}
\begin{align}
H_\mathrm{eff}^\mathrm{LH2}
  &= \hbar\sum_{i,j=1}^{8}
     \Bigl(\Omega_{ij}^{(800)}-\tfrac{i}{2}\,\Gamma_{ij}^{(800)}\Bigr)
     \sigma^{eg}_{i,u}\,\sigma^{ge}_{j,u} \notag\\
  &+ \hbar\sum_{i,j=9}^{24}
     \Bigl(\Omega_{ij}^{(850)}-\tfrac{i}{2}\,\Gamma_{ij}^{(850)}\Bigr)
     \sigma^{eg}_{i,l}\,\sigma^{ge}_{j,l} \notag\\
  &+ \hbar\sum_{i=1}^{8}\sum_{j=9}^{24}
     \Bigl(\Omega_{ij}^{(800)}-\tfrac{i}{2}\,
       \bigl(\tfrac{1}{3}\Gamma_{ij}^{(800)}
            +\tfrac{2}{3}\Gamma_{ij}^{(850)}\bigr)\Bigr)
     \sigma^{eg}_{i,u}\,\sigma^{ge}_{j,l} \notag\\
  &+ \hbar\sum_{i=9}^{24}\sum_{j=1}^{8}
     \Bigl(\Omega_{ij}^{(850)}-\tfrac{i}{2}\,
       \bigl(\tfrac{1}{3}\Gamma_{ij}^{(800)}
            +\tfrac{2}{3}\Gamma_{ij}^{(850)}\bigr)\Bigr)
     \sigma^{eg}_{i,l}\,\sigma^{ge}_{j,u},
\label{eq:S_Hefflh2}
\end{align}
where $\sigma^{ge}_{i,\alpha}=|g_i\rangle\langle e_i|_\alpha$ is the
de-excitation operator and $\sigma^{eg}_{i,\alpha}=(\sigma^{ge}_{i,\alpha})^\dagger$ its adjoint, acting on pigment $i$ in sub-ring $\alpha$. The site states $|g_i\rangle$ and $|e_i\rangle$ are effective two-level reductions of the pigment-protein chromophore. Its axial ligation, carotenoid contact, and local protein micro-environment are coarse-grained into a HOMO-LUMO like pair that reproduces the observed Q-band transition dipole (primarily along the $Q_y$ axis), so that each pigment enters the model as a two-level emitter with transition frequency $\omega_i$ and unit-vector dipole
$\hat{\boldsymbol{\mu}}_i$. where $\Omega_{nm}$, the coherent dipole-dipole coupling, and $\Gamma_{nm}$, the dissipative radiative coupling, are defined in Eq.~(\ref{eq:S_kernels}) of Sec.~S2.
Cross-block elements coupling B800 to B850 sites are evaluated at a
weighted wavenumber $k_\mathrm{mix} = \tfrac{1}{3}k_{800}+\tfrac{2}{3}k_{850}$, where $k_{800}$ and $k_{850}$ are the free-space wavenumbers at 8000~\AA\ and 8500~\AA, respectively.
Numerical values of all parameters are collected in Table~\ref{tab:params}.

\begin{table}[h]
\caption{Parameters of the single-LH2 Hamiltonian.
  Detuning $\delta$ is the B800--B850 site-energy difference.}
\label{tab:params}
\begin{ruledtabular}
\begin{tabular}{lcc}
Parameter & Atomic units & SI / practical \\
\hline
$\lambda_{B800}$ & $1.512\times10^{4}$~bohr  & 8000~\AA             \\
$\lambda_{B850}$ & $1.606\times10^{4}$~bohr  & 8500~\AA             \\
$\delta$         & $-3.350\times10^{-3}$~a.u. & $-91.2$~meV         \\
$|\boldsymbol{\mu}|$  & $1.770$~a.u.         & 4.5~D               \\
\end{tabular}
\end{ruledtabular}
\end{table}

\section*{S2.\;Dyadic Green's Tensor, Coupling Kernels,\\
and Super-Atom Coarse-Graining}
\label{sec:S2}

The free-space dyadic Green's tensor is adopted following 
Lehmberg~\cite{lehmbergRadiationNAtomSystem1970}:
\begin{align}
  G_{ij}(\mathbf{r},k_0) &=
  \frac{e^{ik_0 r}}{4\pi k_0^2 r^3}
  \Bigl[(k_0^2 r^2 + ik_0 r - 1)\,\delta_{ij} \notag\\
  &\qquad - (k_0^2 r^2 + 3ik_0 r - 3)\,\hat{r}_i\hat{r}_j\Bigr],
  \label{eq:S_Green}
\end{align}
where $r=|\mathbf{r}|$ and $\hat{r}_i=r_i/r$; in the near-field limit
$k_0 r\ll 1$ the tensor reduces to $r^{-3}$ scaling, the F\"orster regime,
while in the far-field limit $k_0 r\gg 1$ it reduces to $r^{-1}$ scaling,
the radiation zone.
The coherent and dissipative coupling kernels entering Eq.~(\ref{eq:S_Hefflh2})
are
\begin{align}
  \Omega_{nm} &= -\frac{3\pi\Gamma_0}{k_0}\,
            \mathrm{Re}\!\left[\boldsymbol{\mu}_n^*\cdot
            \mathbf{G}_{nm}\cdot\boldsymbol{\mu}_m\right], \\
  \Gamma_{nm} &= \frac{6\pi\Gamma_0}{k_0}\,
            \mathrm{Im}\!\left[\boldsymbol{\mu}_n^*\cdot
            \mathbf{G}_{nm}\cdot\boldsymbol{\mu}_m\right],
  \label{eq:S_kernels}
\end{align}
where $\mathbf{G}_{nm}\equiv\mathbf{G}(\mathbf{r}_n-\mathbf{r}_m,k_0)$;
the on-site limits $\Omega_{nn}=0$ and $\Gamma_{nn}=\Gamma_0$ are satisfied
by construction, with single-chromophore decay rates
$\Gamma_0^{B800} = 3.00\times10^{-10}$~a.u.\ and
$\Gamma_0^{B850} = 2.49\times10^{-10}$~a.u., obtained from
$\Gamma_0 = \tfrac{4}{3}|\boldsymbol{\mu}|^2 k_0^3$ at their
respective wavenumbers.

For the LH2 unit cell, the decay rate of the brightest collective eigenstate $|S\rangle$ of
$H_\mathrm{eff}^\mathrm{LH2}$, i.e. $\Gamma_1$, is used in place of $\Gamma_o$, $k_0 = \omega_\text{eff}/c_{au}$ and $\mu$ claculated for eqution (~\ref{eq:musa}) were used. 
The computed magnitude is $|\boldsymbol{\mu}_\mathrm{SA}| = 19.47$~a.u.,
the effective transition frequency is
$\omega_0^\mathrm{eff} = 1.603\,458\times10^{-2}$~a.u., and the
effective single-SA decay rate is
$\Gamma_0^\mathrm{eff} = 9.700\,602\times10^{-8}$~a.u.
The brightest eigenstate carries dominant oscillator strength and couples
most strongly to the far-field radiation~\cite{meier1997polarons},
making it the primary optical doorway under low-intensity excitation
conditions relevant to natural photosynthesis.

\section*{S3.\;Near-Field and Far-Field Emission Profiles}
\label{sec:S3}
The superradiant and subradiant character assigned to the single-ring
eigenstates from the imaginary parts of the complex eigenvalues is
confirmed independently by the near- and far-field radiation profiles,
The field intensity at observation point $\mathbf{r}$ for eigenmode $m$ is evaluated as:

\begin{align}
    I_m(\mathbf{r}) = \left\lvert \sum_{j=1}^{N_\mathrm{pig}} 
    c_j^{(m)}\, \mathbf{G}(\mathbf{r} - \mathbf{r}_j,\, k_0) 
    \cdot \boldsymbol{\mu}_j \right\rvert^2,
    \label{eq:field_intensity}
\end{align}
where $\mathbf{G}(\mathbf{r}, k_0)$ is the dyadic Green's tensor of Eq.~(\ref{eq:S_Green}), $c_j^{(m)}$ is the $j$th component of the right eigenvector of $H_{\mathrm{eff}}^{\mathrm{LH2}}$ corresponding to eigenmode $m$, and 
$\boldsymbol{\mu}_j$ is the transition dipole of the $j$th chromophore. 
Near-field profiles are evaluated on $XY$ and $XZ$ planes; the $XY$ observation plane is placed at $z_\mathrm{obs} = z_\mathrm{min} - 30$~bohr, corresponding to $r\ll 1/k_0$. 
It is observed that the superradiant modes carry a peak intensity
approximately $10^3$ times that of the dark modes, consistent with the
decay-rate ratios of Fig.\ref{fig:tilt}a; at $\theta=0^\circ$ the intensity maxima
are $4.16\times10^{-12}$~a.u.\ for modes~2 and~3
($\tilde\Gamma/\Gamma_0^{B800}=757.5$) against
$\sim2\times10^{-15}$~a.u.\ for modes~0 and~1.
A bimodal angular distribution, characteristic of an in-plane coherent
dipole array, is exhibited by the superradiant modes, while a suppressed
diffuse pattern is exhibited by the dark modes.

\begin{figure}[h]
  \centering
  \includegraphics[width=\linewidth]{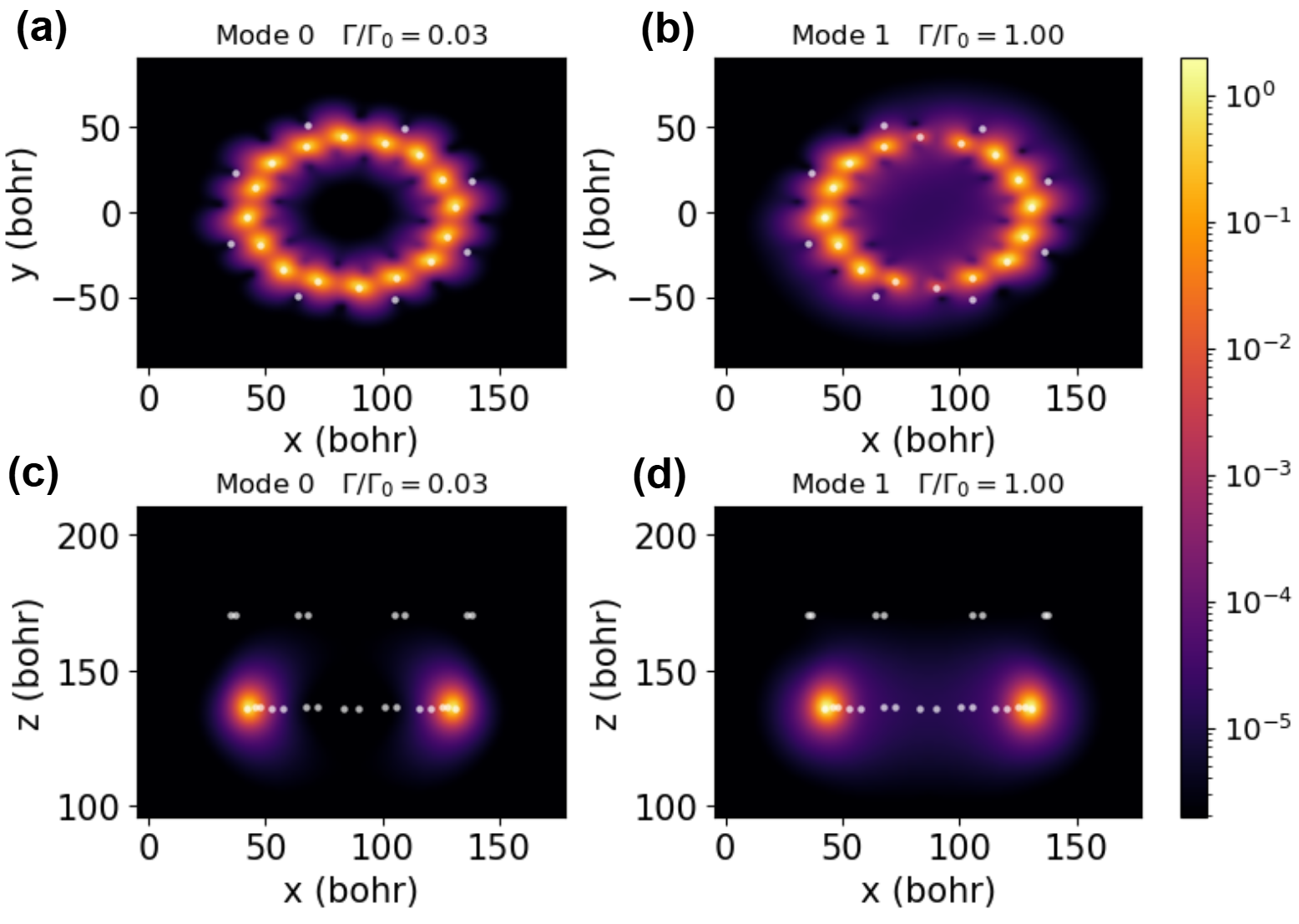}
  \caption{Near-field intensity maps $I_m(\mathbf{r})$ for the near-dark mode~0 ($\Gamma/\Gamma_0^{\mathrm{eff}} = 0.03$, left column) and the dominant superradiant mode~1 ($\Gamma/\Gamma_0^{\mathrm{B800}} = 1.00$, right column) of the single LH2 ring. White dots denote chromophore positions. All maps share a common logarithmic colorbar normalized to the peak intensity of mode~0 in the $XY$ plane. (a,b)~$XY$-plane cross-section. Near- and far-field radiation profiles of the isolated LH2 ring at $\theta=0^\circ$ for a representative superradiant mode     (mode~2, $\tilde\Gamma/\Gamma_0^{B800}=757.5$, top row) and a     representative dark mode (mode~0,     $\Gamma_m/\Gamma_0^{B800}\approx0.02$, bottom row).     \textit{Columns, left to right}: $XY$-plane intensity map; $XZ$-plane intensity map.}
  \label{fig:S_nearfar}
\end{figure}

\section*{S4.\;Supplementary Figures: Longitudinal Slab Geometry}
\label{sec:S5}

\begin{figure}[h]
  \centering
  \includegraphics[width=\linewidth]{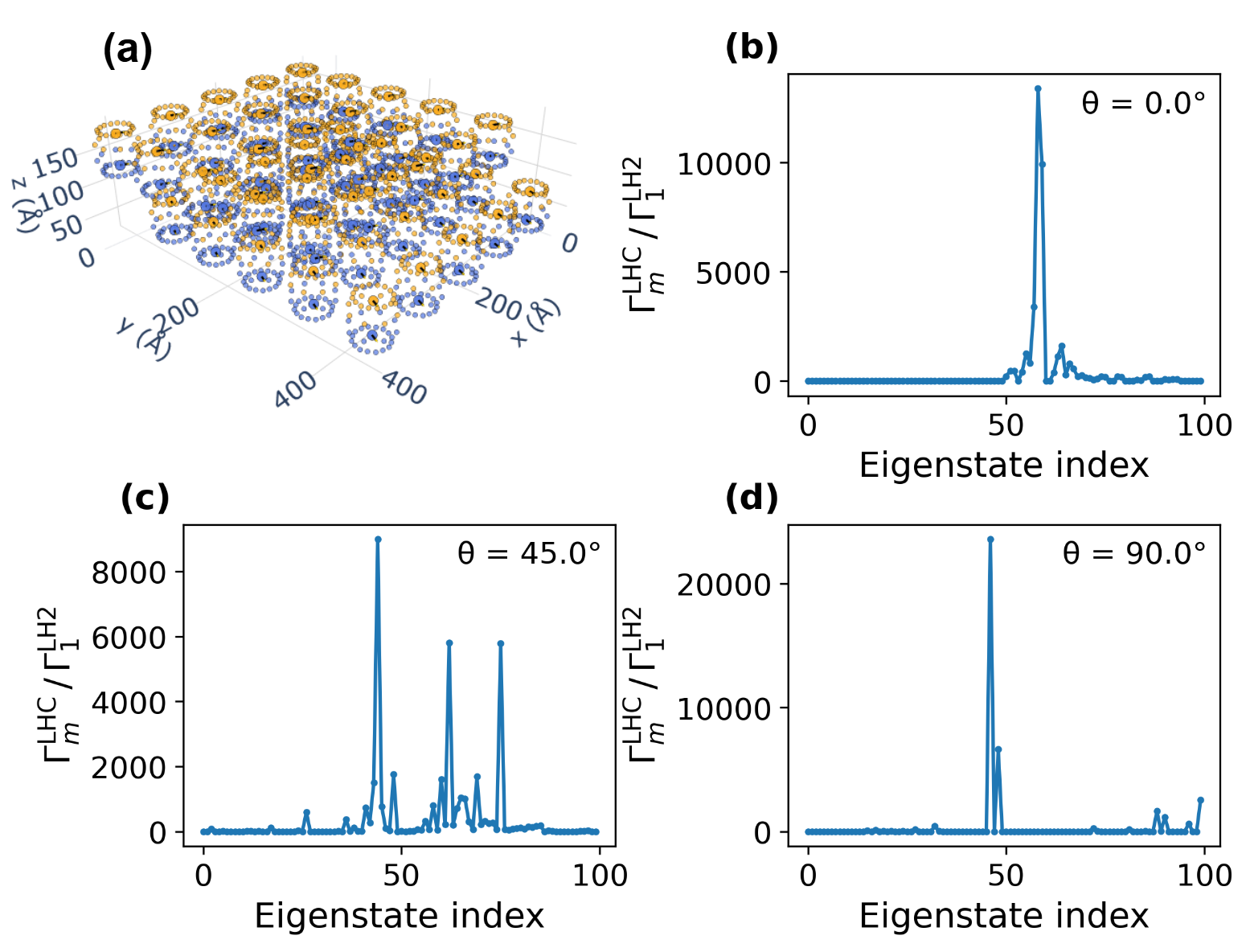}
  \caption{Collective radiative decay-rate spectra  $\tilde\Gamma_m/\Gamma_0^\mathrm{eff}$ of the pristine $5\times5\times1$ LH2 slab (100 super-atoms) at tilt angles $\theta=0^\circ$, $45^\circ$, and $90^\circ$. At $\theta=0^\circ$ two superradiant peaks are concentrated near high eigenstate indices with $\tilde\Gamma_\mathrm{max}/\Gamma_0^\mathrm{eff}\approx2\times10^3$; at $\theta=90^\circ$ the peak rises to $\approx2.3\times10^4$. Bright states remain at high-index modes at all tilt angles.}
  \label{fig:S_551_pristine}
\end{figure}

\begin{figure}[h]
  \centering
  \includegraphics[width=\linewidth]{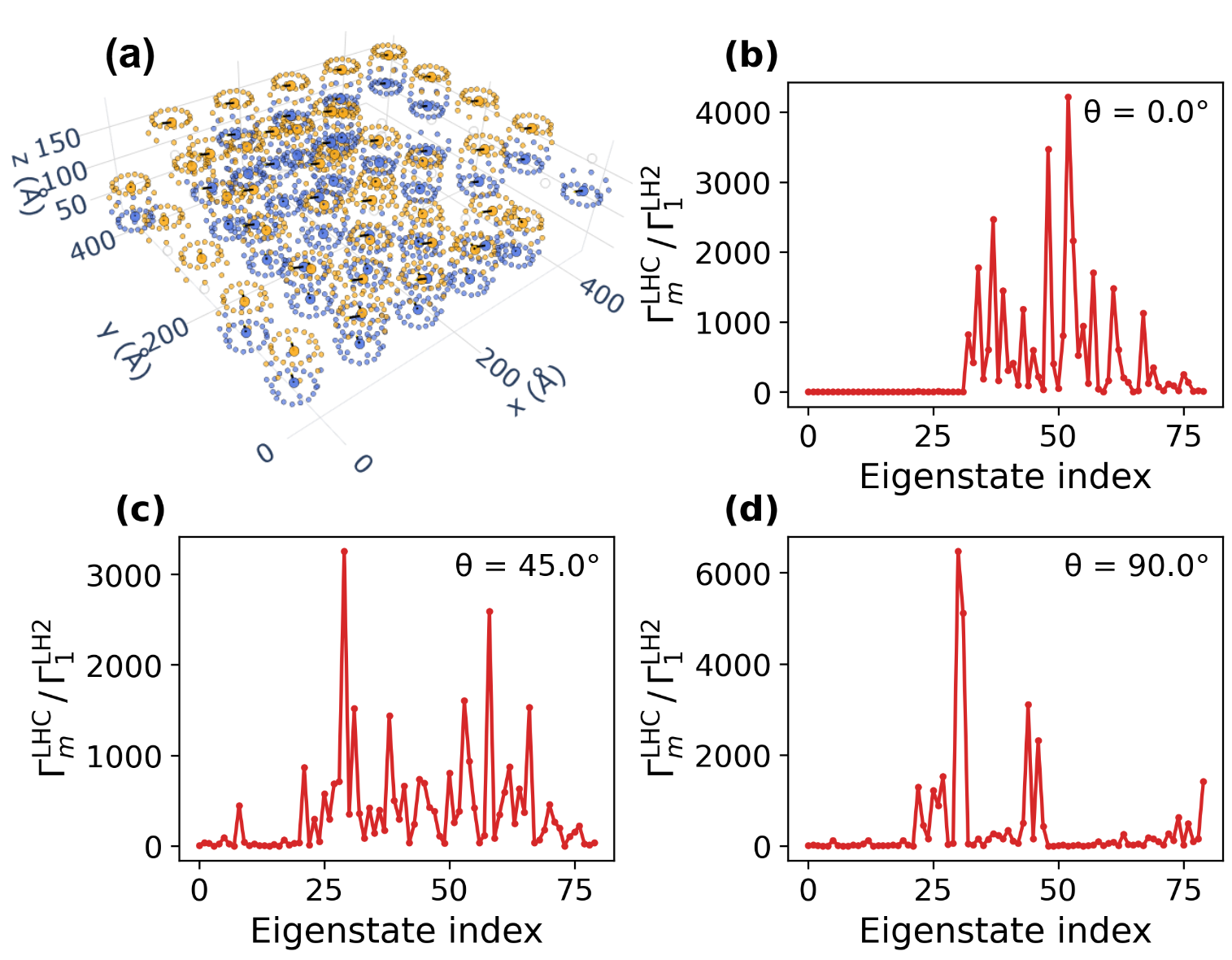}
  \caption{Scaled decay-rate spectra $\tilde\Gamma_m/\Gamma_0^\mathrm{eff}$ of the $5\times5\times1$ slab with 20\% random site vacancy (80~active super-atoms) at $\theta=0^\circ$, $45^\circ$, and $90^\circ$. Spectral broadening is observed at all tilt angles; the peak decay rate is reduced by 65--70\% relative to the pristine slab. Vacancy disorder does not induce migration of superradiant weight toward low-energy modes, and spectral switching is absent in this geometry.}
  \label{fig:S_551_vac}
\end{figure}

\begin{figure}[h]
    \centering
    \includegraphics[width=\linewidth]{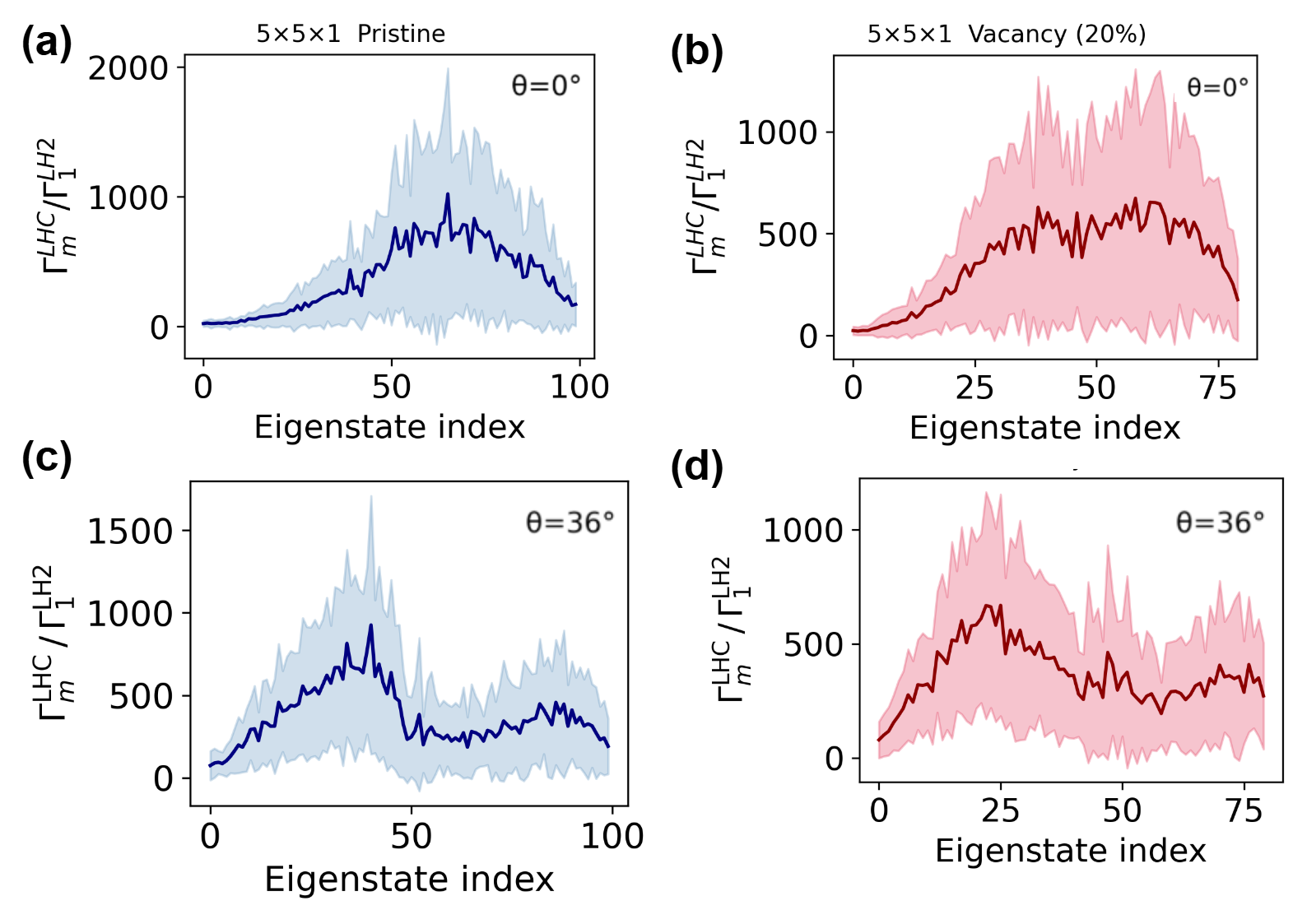}
    \caption{Ensemble-averaged $\Gamma_m^{\text{LHC}}/\Gamma_1^{\text{LH2}}$ decay spectra for a $5\times5\times1$ LH2 slab (longitudinal). Data represent $N=200$ realizations with randomized azimuthal angles $\phi_i \in[0^\circ, 360^\circ)$ per site at fixed $\theta$. Solid lines (shaded bands) denote the mean ($\pm 1\sigma$ interval).(a) Pristine ($\theta = 0^\circ$): Bright sector peaks near index 60 ($\Gamma_{\max} \approx 2000$); variance widens within the bright sector, showing high sensitivity to azimuthal configuration. (b) 20\% Vacancy ($\theta = 0^\circ$): Profile is preserved but peak suppressed to $\approx 750$; increased variance reflects disorder-driven redistribution of oscillator strength. (c) Pristine ($\theta = 36^\circ$): Spectral weight shifts to lower indices (peak $\approx 40$, $\Gamma_{\max} \approx 1500$), consistent with unit-cell bright-dark switching thresholds. (d) 20\% Vacancy ($\theta = 36^\circ$): Migration to lower indices persists ($\Gamma_{\max} \approx 750$); vacancy disorder lowers the orientational brightening threshold without suppressing the switching mechanism.}
    \label{fig:azim_cry_551}
\end{figure}
For completeness, the decay-rate spectra of the $5\times5\times1$ slab geometry are presented here, complementing the $5\times1\times5$ rod spectra of Fig.\ref{fig_3_515} in the main text. In the pristine slab (Fig.~\ref{fig:S_551_pristine}), superradiant peaks are concentrated at high eigenstate indices across all tilt angles; at $\theta=0^\circ$ the peak rate is $\tilde\Gamma_\mathrm{max}/\Gamma_0^\mathrm{eff}\approx2\times10^3$, and at $\theta=90^\circ$ it rises to $\tilde\Gamma_\mathrm{max}/\Gamma_0^\mathrm{eff}\approx2.3\times10^4$ without migration of superradiant weight toward low eigenstate indices. Introduction of 20\% site vacancy (Fig.~\ref{fig:S_551_vac}) broadens the bright sector and reduces the peak decay rate by 65--70\% relative to the pristine slab, without inducing migration of superradiant weight toward low-energy modes at any tilt angle. These results establish that the slab geometry does not support tilt-driven spectral inversion, confirming that switching requires both a transverse carrier orientation and a non-zero out-of-plane dipole component.


\bibliography{References}

\end{document}